\documentclass[reprint,aps,prc,twocolumn,superscriptaddress]{revtex4-2}
\usepackage{amsmath}
\usepackage{amssymb}
\usepackage{nccmath}
\usepackage{bm}
\usepackage{float}
\usepackage{slashed}
\usepackage{comment}
\usepackage{graphicx}
\usepackage{adjustbox}
\usepackage{rotating}
\usepackage{braket}
\usepackage{ulem}
\usepackage{placeins}

\usepackage{subfigure}

\newcommand{\colorcaption}[2][]{%
	\begingroup%
	\renewcommand{\@caption@fignum@sep}{ (Color online). }%
	\caption[#1]{#2}%
	\endgroup%
}

\usepackage{hyperref}
\hypersetup{
	colorlinks=true, 
	linktoc=all,     
	linkcolor=blue,  
	urlcolor=blue,
}

\begin{document}

\title{The $\beta$-decay properties of $N=Z$ nuclei: Role of neutron-proton pairing and the shell model interpretation}

\author{Priyanka Choudhary}
\email{pricho@kth.se}
\author{Chong Qi}
\email{chongq@kth.se}

\address{Department of Physics, KTH Royal Institute of Technology, Roslagstullsbacken 21 SE-106 91, Stockholm, Sweden}

\begin{abstract}
We study the recently measured beta-decay of $^{70}$Kr into $^{70}$Br within the framework of the large-scale shell model. The enhancement in the Gamow-Teller (GT) transition strength in $^{70}$Br compared to the $\beta$-decay of the lighter $^{62}$Ge was suggested as an indication for increased neutron-proton ($np$) pairing correlation. To explore the $np$ correlations in nuclei, we systematically examined the $\beta$-decay properties of the even-even nuclei $A=58,62,66,$ and $70$ into $N=Z$ odd-odd nuclei. By employing an interaction involving solely $J=1, T=0$ and $J=0, T=1$ pairing matrix elements, we observe that the pairing does not necessarily lead to an enhancement in the GT strength for the same coupling strength. But with the inclusion of the $g_{9/2}$ orbital, the GT strength can be increased with increasing $np$ pairing in connection with the enhanced contribution from the $g_{9/2}$ orbital. We further compare those results with realistic calculations in the $fp$ and $f_{5/2}pg_{9/2}$ model space to gauge the contribution from $f_{7/2}$ and $g_{9/2}$ orbitals in the GT strengths. With the  JUN45 interaction, there is an increment for the yrast $1^+$ state for the decay of $^{70}$Kr as compared to the decay of $^{62}$Ge due to increased $g_{9/2}$ contribution. Additionally, we probe the effect of $np$ pairing on $B_{\rm GT}$ by modifying the single-particle energies and the $T = 0$ matrix elements of the interaction responsible for the decay transition strength. In calculations with realistic interaction, we find that the accumulated transition strength can increase with enhanced $np$ pairing.
\end{abstract}


\maketitle



\section{Introduction}
\label{introduction}

The structure and decay properties of neutron-deficient nuclei along the $N=Z$ line have been one of the most important focal points of nuclear physics studies. There have been intensive efforts, especially on nuclei in the $f_{5/2}pg_{9/2}$ space ranging from $^{56}$Ni up to $^{100}$Sn in the past decade. That is the heaviest nuclear region with known spectroscopy for $N=Z$ nuclei \cite{nature}, which are critical for the understanding of the neutron-proton ($np$) correlations \cite{nature,FRAUENDORF201424,qi2015n} as well as nuclear shape evolution~\cite{PhysRevLett.126.072501,PhysRevC.104.L031306}. Those structure effects can affect the $\beta$-decay properties of $N\sim Z$ those nuclei which may be important for the constraint on the rapid proton capture (rp) process in nucleosynthesis \cite{Rubio2017,PhysRevC.79.015803,Zhou2023,Nacher2023-mw,CHEN2024138338,WANG2024138515}. Another interesting feature is the possible manifestation of pseudo-SU(4) symmetry in those nuclei \cite{PhysRevLett.82.2060,sym15112001,PhysRevLett.130.192501}, in connection to the relatively small gap between the $p_{3/2}$ and $f_{5/2}$ orbitals which can be treated as a pseudo-SU(4) partner. Unlike in SU(4) where the Gamow-Teller (GT) transition operator $\sigma\tau$ is a natural generator of the representation, the GT $\beta$ decay can be rather sensitive to the detailed structure coefficients of the pesudo-SU(4) coupling scheme. The $\beta$ decays from $T=1, Z=N\pm2$ nuclei to the $N=Z$, $T=0$, odd-odd nuclei could therefore be a perfect probe of that symmetry \cite{VITEZSVEICZER2022137123}.

An essential feature of nuclear structure that can affect the $\beta$ decay of $N\sim Z$ nuclei is the pairing correlation. Protons and neutrons are filling in the same orbitals for $N = Z$ nuclei. Due to the large overlap between their wave functions, a strong $np$ pair correlation can be expected. The $np$ pair can couple to $J = 0$ and $T = 1$ (isovector) in the same way as a like-particle pairing as indicated by odd-even staggering in binding energies \cite{PhysRevC.61.041303,qi2012double}. On the other hand, in the $T = 0$ (isoscalar) channel, the $np$ pair can couple to different spin values, of which those with $J = J_{\rm max}$ and $J=1$ are often expected to play an important role in various nuclear properties, which is still considered as an open question under intensive debate. The $\beta$-decay to (or from) odd-odd $N=Z$ nuclei provides a unique opportunity to probe those couplings as it connects an isovector to an isoscalar $np$ pair. Along the $N=Z$ line, GT transitions
$^{58}$Zn (0$^+_{\rm g.s.}$) $\rightarrow$  $^{58}$Cu \cite{Fujita2002-no,Jokinen1998-fo,Kankainen2005-id,Ciemny2020-of,Kucuk2017}, $^{62}$Ge (0$^+_{\rm g.s.}$) $\rightarrow$  $^{62}$Ga \cite{PhysRevLett.113.092501,PhysRevC.103.014324}, and $^{70}$Kr (0$^+_{\rm g.s.}$) $\rightarrow$  $^{70}$Br\cite{VITEZSVEICZER2022137123} have been observed.  The measured low-lying GT strengths for the decays of $^{62}$Ge (0$^+_{\rm g.s.}$) $\rightarrow$  $^{62}$Ga are rather small. In the meanwhile, the GT strengths for $^{70}$Kr (0$^+_{\rm g.s.}$) $\rightarrow$  $^{70}$Br are significantly stronger by a few times. It was thus speculated that the $np$ collectivity in $^{62}$Ga must be very weak while it might be increased in $^{70}$Br as the mass number increases~\cite{VITEZSVEICZER2022137123}.  

It is often taken for granted that the isoscalar \(T=0\) pairing
should enhance the GT \(\beta\)-decay rate
between the \(T=1\) ground state of an even-even nucleus and the lowest \(J^{\pi}=1^{+}\)state of the
odd-odd \(N=Z\left(T_{z}=0\right)\) daughter, which therefore could serve as a fingerprint for the onset of isoscalar pairing. For example in Ref. \cite{VITEZSVEICZER2022137123}, it is mentioned that the reason for the enhancement of GT transition strength in $^{70}$Br compared to the lighter one, $^{62}$Ge, might be the increased $np$ correlation in the $T = 0$ channel. That viewpoint has, however, never been carefully examined to our knowledge. References \cite{PhysRevC.60.014311,JANECKE200587} predict an increase of $np$ correlations in the $T = 0$ channel with increasing mass number $A$. The purpose of this work is, therefore, to present a systematic large-scale shell model configuration interaction study on the GT decays to odd-odd $N=Z$ nuclei and identify the role played by the isovector as well as the isoscalar $J=1$ $np$ interactions. We have selected four different transitions to investigate: $^{58}$Zn (0$^+_{\rm g.s.}$) $\rightarrow$  $^{58}$Cu, $^{62}$Ge (0$^+_{\rm g.s.}$) $\rightarrow$  $^{62}$Ga, $^{66}$Se (0$^+_{\rm g.s.}$) $\rightarrow$  $^{66}$As, and $^{70}$Kr (0$^+_{\rm g.s.}$) $\rightarrow$  $^{70}$Br. The realistic shell model calculations are also compared to those with a simpler Hamiltonian with only $T=1$ and $T=0$ pairing terms.

\section{Formalism}
The nuclear shell-model Hamiltonian, which consists of a single-particle energy term and a two-nucleon interaction, is given by
\begin{equation}
H = T + V = \sum_{\alpha}{\epsilon}_{\alpha} c^{\dagger}_{\alpha} c_{\alpha} + \frac{1}{4} \sum_{\alpha\beta \gamma \delta}v_{\alpha \beta \gamma \delta} c^{\dagger}_{\alpha} c^{\dagger}_{\beta} c_{\delta} c_{\gamma},
\end{equation}
where $\alpha = \{n,l,j,t\}$ represents the set of quantum numbers characterizing a single-particle state within a given model space, the corresponding single-particle energy is given by $\epsilon_{\alpha}$, and $v_{\alpha \beta \gamma \delta} = \langle\alpha \beta | V | \gamma \delta\rangle $ are the antisymmetrized two-body matrix elements. The operators $c^{\dagger}_{\alpha}$ and $c_{\alpha}$ are the creation and annihilation operators, respectively. With the effective Hamiltonian in the above format, the eigen-value and eigen-functions can be conveniently, though very often computationally heavily, calculated via the full configuration interaction shell model approach. 

The GT reduced transition probability $B_{\rm GT}$ is defined as
\begin{equation}
\hspace{-0.2cm}B_{\rm GT} =\frac{g_{\rm A}^{2}}{2J_{i}+1}|{M}_{\rm GT}|^2,
\end{equation}
where $g_{\rm A}$ is the axial-vector coupling constant, $J_{i}$ is the angular momentum of the initial state and ${M}_{\rm GT}$ is the Gamow-Teller nuclear matrix element (NME), expressed as $M_{\rm GT} = {\langle {\Psi_f}|| \sum_{k}{\sigma^k\tau_{\pm}^k} ||\Psi_i \rangle}.$  Here, $\bm{\sigma}$ denotes the spin operator, $\tau_{\pm}$ are the isospin raising and lowering operators $\tau_+|p\rangle = |n\rangle$, $\tau_-|n\rangle  = |p\rangle$, and the sum runs over the single-particle orbitals. $|\Psi_i \rangle$ and $|\Psi_f \rangle$ are the wave functions of the parent and daughter nuclei, respectively. In the present calculations, we have also considered an overall quenching of the GT reduced probability by the factor $q$ = 0.79.

The NME contains the nuclear structural details and is a product of the single-particle matrix elements (SPMEs) and the one-body transition densities (OBTDs) between the initial $(\Psi_i)$ and final $(\Psi_f)$ states. The SPMEs are independent of the effective Hamiltonian and their expression can be found in the Supplemental Material \cite{SM}. In our case, we have evaluated the wave functions and OBTDs using the shell-model codes NuShellX \cite{nushellx,nushellx-kth} and, partly, KSHELL \cite{SHIMIZU2019372,shimizu2013nuclearshellmodelcodemassive}.
\begin{figure}
	\centering
	\includegraphics[width=\columnwidth]{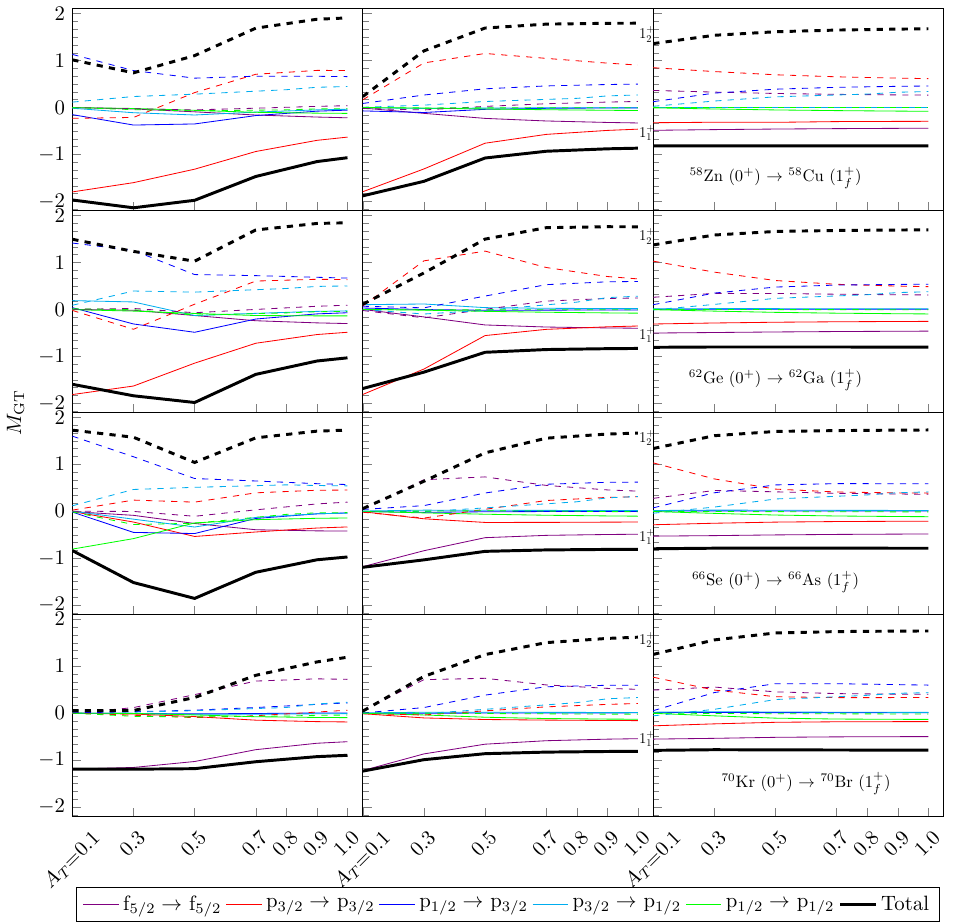}
	\caption{The solid lines indicate the transition from the ground state to the $1_1^+$ state, while the dashed lines are for the $1_2^+$ state. For the single-particle energies we have taken equally spaced $\epsilon_{p_{3/2}}$ = 0 MeV, $\epsilon_{p_{1/2}}$ = 1 MeV, and $\epsilon_{f_{5/2}}$ = 2 MeV for the calculations in the left panel. For the middle panel, the single-particle energies are $\epsilon_{p_{3/2}}$ = 0 MeV, $\epsilon_{p_{1/2}}$ = 2 MeV, and $\epsilon_{f_{5/2}}$ = 1 MeV. For the right panel, we assume $p_{3/2}$ and $f_{5/2}$ orbitals to be degenerate, which favors pesudo-SU(4) symmetry, and $p_{1/2}$ lies higher at 1 MeV.}
	\label{SU4}
\end{figure}
\section{Results and discussion}

The shell model diagonalization approach provides a convenient tool to study the competition between various isoscalar and isovector pairing interactions \cite{PhysRevC.84.044318,Pan2020,qi2011spin,PhysRevC.86.041302}. In this work, we begin by performing schematic calculations in the model space $p_{3/2}p_{1/2}f_{5/2}$ considering both the isovector and isoscalar pairing matrix elements in three extreme setups for the single-particle level scheme: setup I with the order of the orbitals $p_{3/2},p_{1/2},f_{5/2}$; setup II with the order  $p_{3/2},f_{5/2},p_{1/2}$ and setup III which incorporates pseudo-SU (4) symmetry. The pairing interaction is evaluated with the surface $\delta$ effective interaction where we assumed the same coupling strength $A_T$ for both pairing channels for simplicity. Only $J=0$, $T=1$ and $J=1$, $T=0$ two-body matrix elements are considered. The expression for the surface $\delta$ interaction (SDI) we have used is derived in the first part of the Supplemental Material \cite{SM} (see also Refs. \cite{brussaard1977shell,talmi1993simple} therein).

The results for the GT transition strengths of four selected different transitions: $^{58}$Zn (0$^+_{\rm g.s.}$) $\rightarrow$  $^{58}$Cu, $^{62}$Ge (0$^+_{\rm g.s.}$) $\rightarrow$  $^{62}$Ga, $^{66}$Se (0$^+_{\rm g.s.}$) $\rightarrow$  $^{66}$As, and $^{70}$Kr (0$^+_{\rm g.s.}$) $\rightarrow$  $^{70}$Br are plotted in Fig. \ref{SU4} as a function of $A_T$.
In Fig. \ref{SU4}, we have included the separate contributions from different single-particle transitions as well as the total $M_{\rm GT}$ values for decay to the lowest two $1^+$ states. Detailed results are listed in Tables I-XII in the Supplemental Material \cite{SM}. This figure illustrates the effect of the isoscalar and isovector strengths on the individual GT matrix elements. One conclusion one can draw immediately is that it is not generally true that the pairing correlation can enhance the GT decay strength, even for cases like transitions between the yrast $0^+$ and $1^+$ states where major contributions of different single-particle transitions have the same signs and are constructive. That is partly due to the fact that the GT transitions are highly selective and only connect states with the same $l$ value.

The GT transitions for
$^{58}$Zn (0$^+_{\rm g.s.}$) $\rightarrow$  $^{58}$Cu have been measured in Refs. \cite{Fujita2002-no,Jokinen1998-fo,Kankainen2005-id,Ciemny2020-of,Kucuk2017}, where one notices a reasonable agreement for the measured decay strength to the first $1^+$ state but large discrepancy for $B_{\rm GT}(0^+_{\rm g.s.}\rightarrow 1^+_2)$. 
For that transition, involving two particles relative to the $N=Z=28$ core, the dominant contribution comes from $p_{3/2}$ $\rightarrow$ $p_{3/2}$ in the left and middle panels. With a stronger $np$ pairing, the occupancy in $p_{3/2}$ decreases for $^{58}$Cu which leads to a decrement in the magnitude of the strength from $p_{3/2}$ to $p_{3/2}$. With pseudo-SU(4) symmetry considered in the right panel, the transition from $p_{3/2}$ $\rightarrow$ $p_{3/2}$, $f_{5/2}$ $\rightarrow$ $f_{5/2}$ and $p_{1/2}$ $\rightarrow$ $p_{1/2}$ are in same phase, which resulted in a $|M_{\rm GT}|$ value of around 0.8 for all pairing strengths considered. That is because the first $1^+$ state is dominated by the configuration $(p_{3/2}f_{5/2})^{1^+}$ with mixtures from $f_{5/2}^2$ and $p_{3/2}^2$.

\begin{figure}[htbp]
	\centering
	\begin{subfigure}
		\centering
		\includegraphics[width=0.5\textwidth]{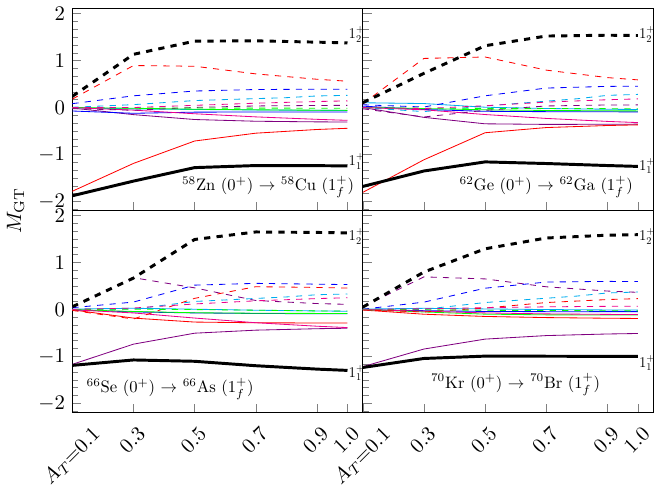}
		\put(-260,180){\text{(a)}}
	\end{subfigure}
	\vskip\baselineskip
	\begin{subfigure}
		\centering
		\includegraphics[width=0.5\textwidth]{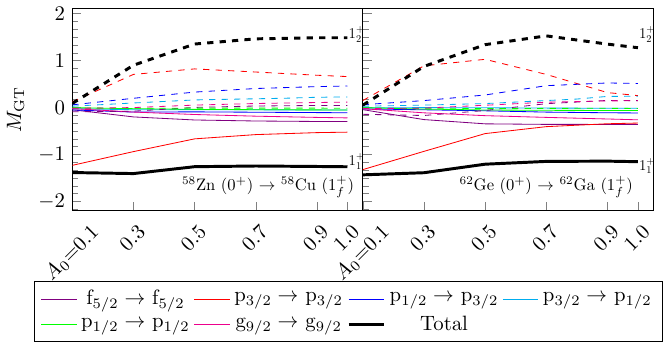} 
		\put(-260,120){\text{(b)}}
	\end{subfigure}
	\caption{The calculated reduced GT transition matrix elements (a) as a function of $A_T$ (the same coupling strength for both pairing channels) and  (b) as a function of $A_0$ for the fixed $A_1 = 0.6$ in the $f_{5/2}pg_{9/2}$ model space.}
	\label{SDI_5pg9}
\end{figure}

$^{62}$Ge (0$^+_{\rm g.s.}$) $\rightarrow$  $^{62}$Ga ($1^+_1$) and $^{70}$Kr (0$^+_{\rm g.s.}$) $\rightarrow$  $^{70}$Br  ($1^+_1$) are dominated by the $p_{3/2}$ $\rightarrow$ $p_{3/2}$ and $f_{5/2}$ $\rightarrow$ $f_{5/2}$ contributions, respectively, in the left and middle panels. That is related to the fact that $^{62}$Ga ($1^+_1$) is dominated by the configuration $(p_{3/2})^{-2}$ while $^{70}$Br($1^+_1$) is dominated by $(f_{5/2})^{2}$. The major contribution for $^{66}$Se (0$^+_{\rm g.s.}$) $\rightarrow$  $^{66}$As ($1^+_1$) comes from $p_{1/2}$ $\rightarrow$ $p_{1/2}$ for the setup I, while it is from $f_{5/2}$ $\rightarrow$ $f_{5/2}$ for setup II because of the different ordering of the orbitals in these two setups. For calculations assuming $f_{5/2}$ and $p_{3/2}$ degeneracy on the right-hand side of the figure, one notices a similar behavior in decays to $^{62}$Ge, $^{66}$Se, and $^{70}$Kr except the fact the contribution from $p_{1/2}$ increases as the number of pairs increases. Similar calculations assuming $f_{5/2}$ and $p_{3/2}$ degeneracy were also performed in Refs. \cite{sym15112001,VITEZSVEICZER2022137123} with surface $\delta$ and $\delta$ interactions, respectively, focusing on the effect of the deformation on the accumulated $B_{\rm GT}$ strength.

\begin{table*}
	\centering
	\caption{GT strengths for the $\beta$-decay $^{62}$Ge($0^+$) $\rightarrow$ $^{62}$Ga($1_f^+$) using the JUN45 interaction with varying the single-particle energies along with the experimental data \cite{PhysRevC.103.014324}.}
	\label{Table1}
	\begin{tabular}{lccccccc}
		\hline \hline
		$J_f^{\pi}$ & Expt & JUN45 & ($\delta=100$) & ($\delta=200$) & ($\delta=300$) & ($\delta=400$) & ($\delta=500$)\\
		\hline
		$1^+_1$ & 0.068(6) & 0.0621 & 0.0841 &  0.1465 & 0.2547 & 0.3754 & 0.4677\\
		$1^+_2$ & 0.047(4) & 2.3100 & 1.9200 & 1.5670 & 1.2860 & 1.0740 & 0.9370 \\
		
		\hline\hline
	\end{tabular}
\end{table*}
Now, we have extended our model space by including the $g_{9/2}$ orbital and we have performed the schematic calculations. We have not done schematic calculations for the $fp$ model space due to the huge dimension of the calculations. Results corresponding to $f_{5/2}pg_{9/2}$ are shown in Fig. \ref{SDI_5pg9}. We have taken the same ordering of the orbitals as considered in the JUN45 interaction, to make a reasonable comparison of the schematic calculations with the realistic one. We have performed the full space calculations for the decay of $^{58}$Zn, $^{62}$Ge, and $^{66}$Se, while for the decay of $^{70}$Kr, we have allowed a maximum of four nucleons (two protons and two neutrons ) in the $g_{9/2}$ orbitals from the  $f_{5/2}p$ space. We can see from the figure that for the $1^+_1$ state, $B_{\rm GT}$ value first decreases with increasing $np$ pairing strength. After a certain strength, it starts increasing. 
	
One can see that in the case of $^{58}$Zn $\rightarrow$ $^{58}$Cu, the contribution from $p_{3/2} \rightarrow p_{3/2}$ decreases and the contributions from $f_{5/2} \rightarrow f_{5/2}$, $p_{1/2} \rightarrow p_{1/2}$ and $g_{9/2} \rightarrow g_{9/2}$ increase  with increasing $np$ pairing strength. All the contributions are found to have the same phase as found in Fig. \ref{SU4}. For $^{62}$Ge $\rightarrow$ $^{62}$Ga, we can observe that the occupancy of $p_{3/2}$ decreases, while the occupancy of $g_{9/2}$ increases. The $^{66}$Se $\rightarrow$ $^{66}$As and $^{70}$Kr $\rightarrow$ $^{70}$Br transitions are dominated by contributions of the $f_{5/2} \rightarrow f_{5/2}$ transition. With stronger $np$ pairing, $p_{1/2} \rightarrow p_{1/2}$, $p_{3/2} \rightarrow p_{3/2}$ and $g_{9/2} \rightarrow g_{9/2}$ contributions increase.  The major change in the $f_{5/2}pg_{9/2}$ model space calculations is due to $g_{9/2}$ orbital as compared to the schematic calculation with the $f_{5/2}p$ model space.\\

We did not intend to optimize the single-particle energies and pairing strengths in the above calculations to better reproduce the experimental data. Instead, we focus on the sign of the different components of the wave functions and the separate contributions from different one-body transitions. As a comparison, we have performed shell-model calculations for all the above transitions with realistic shell-model Hamiltonians that were constructed to well reproduce the spectra and other properties of the nuclei involved.
We have used JUN45 for $f5pg9$-model space with $^{56}$Ni as inert core \cite{PhysRevC.80.064323} and GXPF1J interactions for $fp$-model space \cite{M_Honma_2005}. The detailed results of GT-strengths  for the above transitions calculated with the realistic shell model interactions JUN45 and GXPF1J are listed in the last part of the Supplemental Material \cite{SM}.

\subsection{The $\beta$ decay of $^{58}$Zn (0$^+_{\rm g.s.}$) $\rightarrow$  $^{58}$Cu ($1^+_f$)}
In the case of the $\beta$-decay of $^{58}$Zn, our calculations using the JUN45 yield a large $B_{\rm GT}$ value of 4.1577 for the 1$^+_{1}$ of $^{58}$Cu. The results are similar to those obtained from the schematic interaction SDI in $f_{5/2}p$ and $f_{5/2}pg_{9/2}$ model spaces with only pairing terms. The $g_{9/2}$ orbital in the JUN45 interaction has a negligible effect on the total $B_{\rm GT}$. On the other hand, with increased $np$ pairing strength in the schematic interaction for the $f_{5/2}pg_{9/2}$ model space, the contribution from $g_{9/2}$ increases. The $B_{\rm GT}$ value is highly sensitive to the $np$ pairing strength. However, the GXPF1J results reveal a significant role of the $f_{7/2}$ orbital, whose contribution is in the opposite phase and thus destructive.  With the $f_{7/2}$ orbital fully occupied, the $B_{\rm GT}$ value is 2.3037, while the full $fp$-shell calculation reduces the $B_{\rm GT}$ value to 0.2773 which is close to the experimental data \cite{Jokinen1998-fo,Kucuk2017}. For the $1^+_2$ state, the calculated $B_{\rm GT}$ value of 0.2842 is within the error range of the experimental value 0.54(26) \cite{Jokinen1998-fo}, which reflects an improvement compared to the previous theoretical calculations \cite{Jokinen1998-fo} using the FPD6 interaction \cite{RICHTER1991325,RICHTER1994585}.

\begin{figure}
    \centering
    \includegraphics{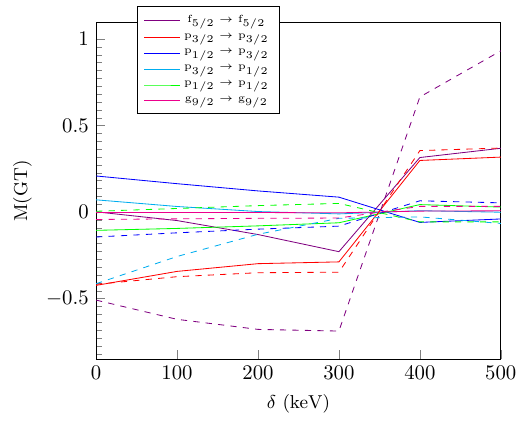}
    \caption{GT matrix elements for each configuration and their total as a function of the energy shift.}
    \label{fig:SPE}
\end{figure}

\subsection{The $\beta$ decay of $^{62}$Ge($0^+$) $\rightarrow$ $^{62}$Ga($1_f^+$)}
We have done the calculations for the $\beta$-decay transition from $^{62}$Ge (0$^+_{\rm g.s.}$) to $^{62}$Ga ($1^+_f$). In Ref. \cite{KUMAR2022122344}, the study using various shell model interactions seems to give very different results for which the reason is unknown. In this work, we have also implemented the JUN45 interaction \cite{PhysRevC.80.064323} and extended the analysis with the GXPF1J interaction \cite{M_Honma_2005}, a modified version of GXPF1A \cite{Honma_2005} (The modification involved scaling the multipole part of the two-body matrix elements $V(f_{7/2}f_{5/2}f_{7/2}f_{5/2}; J1)$ by a factor of 0.7 for all $J$). We noted that the JUN45 interaction predicts a $B_{\rm GT}$ value of 0.0621 that is in good agreement with the experimental value [0.068(6)] for the $1^+_1$ state. For the $1^+_2$ state, we obtain a large $B_{\rm GT}$ value of 2.31, which is again attributed due to the same phase contribution from all major configurations. The contribution from $g_{9/2}$ is found to be minimal in these transitions. With the GXPF1J interaction, the contributions from $f_{5/2}$ $\rightarrow$  $f_{7/2}$ and $f_{7/2}$ $\rightarrow$ $f_{5/2}$ are in opposite phases which results in a reduction of the total $B_{\rm GT}$ value to 0.1936. We have also carried out the full shell model calculations with the KB3G interaction \cite{POVES2001157} with the same quenching factor. In Ref. \cite{PhysRevLett.113.092501}, the calculation for the $\beta$ decay of $^{62}$Ge into $^{62}$Ga has been done with the same interaction, allowing up to five nucleons to be excited from the $f_{7/2}$ shell to the rest of the $pf$ orbitals. The quenching factor $(g_A/g_V)_{\rm eff} = 0.74(g_A/g_V)_{\rm free}$ has been applied in the calculations. Further, in Ref. \cite{KUMAR2022122344}, a minimum of six protons or neutrons and a minimum of two protons or neutrons in $f_{7/2}$ and $p_{3/2}$ orbitals, respectively, are fixed and the quenching factor $q$ = 0.660 $\pm$ 0.016 is used. It is concluded in Ref. \cite{KUMAR2022122344} that summed B(GT) using GXPF1A interaction are better than the values obtained using the KB3G interaction. In our calculation, for the first $1^+$ state, the obtained value is 0.0722, which is close to the experimental data. However, we found a better agreement for the overall GT transition strength with the experiment using the GXPF1J interaction rather than the KB3G interaction.

To investigate the impact of the pseudo-SU(4) symmetry, we have done calculations with varying single-particle energies of $p_{3/2}$ and $f_{5/2}$ orbitals for the GT transition strength. The calculated $B_{\rm GT}$ values with the JUN45 interaction are shown in Table \ref{Table1}. 
In Fig. \ref{fig:SPE}, the individual matrix elements are presented as a function of $\delta$, where the energy of $p_{3/2}$ is moved up by $\delta$ and the energy of $f_{5/2}$ is pushed down by $\delta$, effectively reducing their energy gap by 2$\delta$. 
As the gap between $p_{3/2}$ and $f_{5/2}$ orbitals decreases, the $B_{\rm GT}$ value for $1_1^+$ diverges from the experimental data, primarily due to the increased contribution from the matrix element of $f_{5/2}$ $\rightarrow$ $f_{5/2}$. This increase in the $B_{\rm GT}$ value is directly related to the enhanced occupancy of the $f_{5/2}$ orbital, which rises from 0.5053 to 1.0511 in $^{62}$Ga, while $p_{3/2}$ occupancy decreases from 1.9721 to 1.4132. For the $1_2^+$ state, the large GT strength predicted by the JUN45 interaction shows improvement as the energy gap reduces. This is due to the decreased contribution from $p_{1/2} \rightarrow p_{3/2}$ and $p_{3/2} \rightarrow p_{1/2}$, where the latter decreases more rapidly with $\delta$, eventually changing its phase which leads to a reduction in the total $M_{\rm GT}$. 
With a 1 MeV reduced gap between those single-particle states [effectively the pseudo-SU(4) symmetry], the calculated $M_{\rm GT}$ value for $1_2^+$ is 0.9680. The major contributions are from $p_{3/2}$ $\rightarrow$ $p_{3/2}$ and $f_{5/2}$ $\rightarrow$ $f_{5/2}$, which contribute in the same phase. The result is the same as obtained with the interaction having only pairing matrix elements (Fig.\ref{SU4}). This suggests that the contribution of the $g_{9/2}$ orbital is minimal.

\begin{figure}
	\includegraphics[width=8cm]{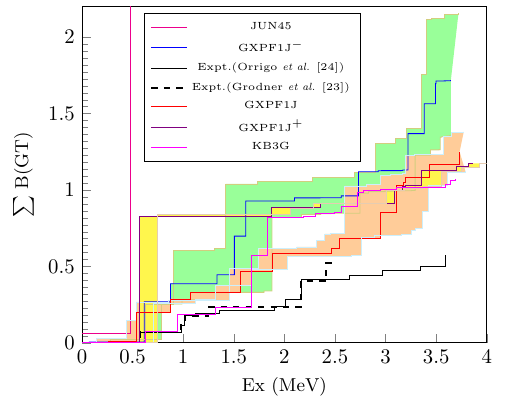}
	\caption{Comparison of cumulative GT strengths for the decay of $^{62}$Ge (0$^+_{\rm g.s.}$) to $^{62}$Ga between the experimental data \cite{,PhysRevLett.113.092501,PhysRevC.103.014324} and shell-model results obtained with the JUN45, GXPF1J, and KB3G interactions.}
	\label{GT_Ga}
\end{figure}

\begin{table}[h]
	\renewcommand{\arraystretch}{1.0}\tabcolsep 0.3 cm
	\centering
	\setlength\tabcolsep{10.0pt}
		\caption{Excitation energies of $^{66}$As($1_f$) and GT strengths using the JUN45 and GXPF1J interactions. $E_{\rm x}$ [$1_1^+$] = 0.837 MeV \cite{GRZYWACZ1998247}.}
	\label{Table2}
	\begin{tabular}{lcccc}
		\hline \hline
  & \multicolumn{2}{c}{$E_{\rm x}$ (MeV)} & \multicolumn{2}{c}{$B_{\rm GT}$} \\
  \cline{2-3}
  \cline{4-5}
		$J_f^{\pi}$  & JUN45 & GXPF1J & JUN45 & GXPF1J\\
		\hline
		$1^+_1$ & 0.611 &  0.023 & 0.1637 &  0.0045\\
		$1^+_2$&  0.762 & 0.450 &  0.3239 & 0.0667\\
		$1^+_3$  & 1.018 & 0.569 & 1.5120 & 0.0053\\
		$1^+_4$  & 1.552 & 0.895 & 0.0021 & 0.0987\\
		$1^+_5$  & 1.720 & 0.969 & 0.0039 & 0.0001\\
		$1^+_6$  & 1.996 & 1.092 & 0.3403 & 0.0049\\
		$1^+_7$ & 2.074 & 1.446 & 0.0177 & 0.5882\\
		$1^+_8$ & 2.275  & 1.695 & 0.0228  & 0.1195\\
		$1^+_9$ & 2.554 &  1.988 & 0.0003 &  0.0040\\
		\hline\hline
	\end{tabular}
\end{table}
In addition, we explored the GXPF1J interaction by varying single-particle energies and pairing matrix elements and their results are plotted in Fig. \ref{GT_Ga}. Specifically, the single-particle energies of orbitals $p_{3/2}$ and $f_{5/2}$ are varied by 500 keV. The interaction with a reduced gap between these two orbitals is labeled as GXPF1J$^-$, whereas that with an increased gap is referred to as GXPF1J$^+$. One can see from Table XX in the Supplemental Material \cite{SM} that the contribution from $p_{1/2} \rightarrow p_{3/2}$ and $p_{3/2} \rightarrow p_{1/2}$ decreases in the case of the GXPF1J$^-$ interaction. This is similar to what is obtained for the JUN45 interaction with a reduced gap. For the GXPF1J$^-$ interaction, the evaluated $M_{\rm GT}$ value is 0.9893,  neglecting the $f_{7/2}$ contributions. This value is close to the one obtained for the JUN45 interaction with pseudo-SU(4) symmetry. If we include the contributions from the $f_{7/2}$ orbital, the total GT-matrix element is reduced due to the opposite phase contributions. It is obvious that when the gap increases, the GXPF1J$^+$ interaction shows an increased contribution from  $p_{3/2} \rightarrow p_{3/2}$. Furthermore, we have modified the $J= 1, T=0$ $np$ pairing matrix elements by adding and subtracting $\Delta$ with a scaling factor of 0.1. The shaded area in Fig. \ref{GT_Ga} illustrates the corresponding results. The green and yellow shaded areas in Fig. \ref{GT_Ga} show the effect of adjusting the $T=0$ matrix elements on top of the single-particle energy variation around the actual GXPF1J$^-$ and GXPF1J$^+$ results, respectively. We have checked the effect of $T=0$ $np$ pairing matrix elements on $1^+_1$ state, it seems that the $B_{\rm GT}$ increases with increased $np$ pairing matrix elements sometimes and sometimes decreases, depending on the interaction. It is hard to make a conclusion based on these results. However, we notice that when we increase the $T=0$ $np$ pairing matrix elements, the total $B_{GT}$ transition strength increases for the GXPF1J interaction. With reduced $np$ pairing matrix elements, it decreases. For the other two cases, in which the single-particle energies of $p_{3/2}$ and $f_{5/2}$ orbitals are varied, we found similar behavior.

\subsection{The $\beta$ decay of $^{66}$Se($0^+$) $\rightarrow$ $^{66}$As($1_f^+$)}
\label{66}
We have also carried out shell model calculation for the $\beta$-decay transition from $^{66}$Se (0$^+_{\rm g.s.}$) to $^{66}$As ($1^+_f$) for a complete study. There are no experimental data available for these transitions yet, this theoretical study provides crucial insights and would be helpful in the future. 
From the experiment \cite{GRZYWACZ1998247}, the only known excitation energy for $^{66}$As is for the $1^+_1$ state, measured as 0.837 MeV. Using the JUN45 and GXPF1J interactions, we provide predictions for the excitation energies of the low-lying states of $^{66}$As, which are tabulated in Table \ref{Table2}. Apart from the energy spectra, we predict the GT-transition strengths for the transition from $^{66}$Se (0$^+_{\rm g.s.}$) to some excited $1^+$ states of $^{66}$As. We notice that all the contributions are in the same phase, except for $p_{1/2} \rightarrow p_{3/2}$ for the first $1^+$ state with JUN45 interaction. For the first $1^+$ state, the JUN45 interaction predicts occupancies of 2.4711 in $p_{3/2}$, 0.6673 in $p_{1/2}$, 1.6062 in $f_{5/2}$, and 0.2559 in $g_{9/2}$, while the GXPF1J interaction exhibits 7.8597 in $f_{7/2}$, 3.5956 in $p_{3/2}$, 1.0354 in $p_{1/2}$, and 0.5131 in $f_{5/2}$. 

\begin{figure}
	\centering
	\includegraphics[width=\columnwidth,height=7cm]{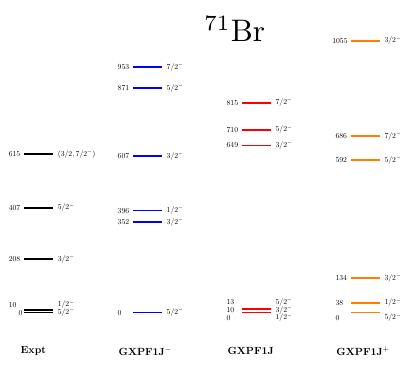}
	\caption{Comparison between experimental \cite{PhysRevC.72.024321} and theoretical low-lying energy states for $^{71}$Br with different GXPF1J interactions.}
	\label{Fig3}
\end{figure}

We have checked the effect of varying the single-particle energies of the GXPF1J interaction on the energy spectra of $^{71}$Br. In Fig. \ref{Fig3}, we illustrate the energy level scheme of $^{71}$Br using the GXPF1J interaction, where the experimental spin-parity assignment for the excited state is tentative \cite{PhysRevC.72.024321}. Notably, the small energy difference of 10 keV between $5/2^-$ and $1/2^-$ suggests nearly degenerate states. With the GXPF1J interaction, $1/2^-$, $3/2^-$, and $5/2^-$ states appear almost degenerate. We also include the energy spectra of GXPF1J$^-$ and GXPF1J$^+$ interactions in Fig. \ref{Fig3}, showing that transitioning from GXPF1J$^-$ to GXPF1J to GXPF1J$^+$ increases occupancy of the $p_{3/2}$ orbital and decreases that of the $f_{5/2}$ orbital. Because the $p_{3/2}$ orbital is moved down by 1 MeV from left to right and the $f_{5/2}$ orbital is moved up by 1 MeV, the ordering of the orbitals is still maintained. 

\subsection{The $\beta$ decay of $^{70}$Kr($0^+$) $\rightarrow$ $^{70}$Br($1_f^+$)}
\begin{figure*}[htp]
	\centering
	\includegraphics[width=14cm]{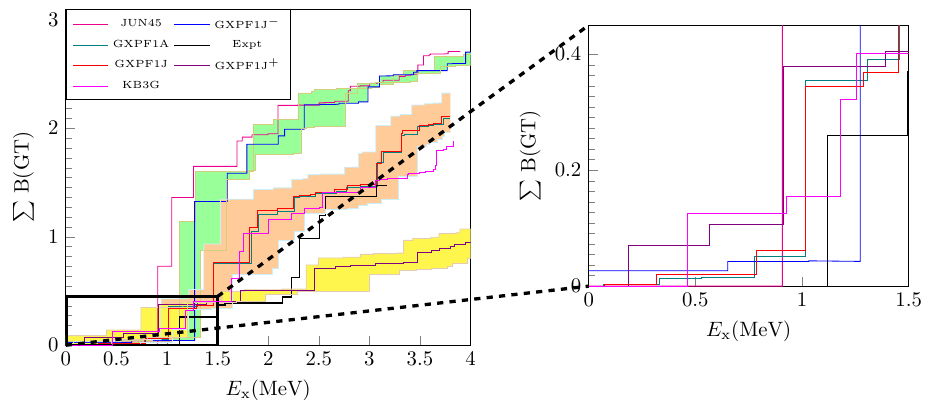}
	\caption{Distribution of GT strengths in the $^{70}$Kr $\rightarrow$ $^{70}$Br decay corresponding to different shell model interactions in comparison with the experimental data  \cite{VITEZSVEICZER2022137123}. The shaded area shows the change in $B_{\rm GT}$ with varying the $T=0$ matrix elements for the fixed single-particle energies of the GXPF1J interaction.}
	\label{GT_Br}
\end{figure*}

Recently, the first measurement of $B(E2; 0^+ \rightarrow 2^+)$ for $^{70}$Kr, determined from an inelastic scattering experiment, shows a shape change between the mirror nuclei $^{70}$Kr and $^{70}$Se \cite{PhysRevLett.126.072501}. The theoretical state-of-the-art shell model calculations with several interactions have been performed for the $A = 70$ triplet and none of the calculations is able to describe the reason behind the strong increase in $B(E2)$ between these mirror nuclei, as observed experimentally. The $B(E2)$ transition strength corresponding to JUN45 and GXPF1A can be found in Ref. \cite{PhysRevLett.126.072501}. The B(E2; $0^+_{\rm g.s.} \rightarrow 2^+_1$) transition strength using the GXPF1J interaction is calculated in this work and the obtained values for $^{70}$Kr and $^{70}$Se are 1878 and 2040 ${\rm e}^2 \, {\rm fm}^4$. We also note a decreasing trend compared to the experimental trend, similar to the GXPF1A results \cite{PhysRevLett.126.072501}. The previously performed large-scale shell model calculations with the JUN45 and JUN45+LNPS plus Coulomb interactions suggested an alternative to the shape change \cite{PhysRevC.104.L031306}. Depending on the effective charges, the results are in line with the experimental data within uncertainties, or the anomaly occurs in $^{70}$Br, rather than in $^{70}$Kr. The decay of $^{70}$Kr seems interesting from the perspective of nuclear deformation and shape coexistence. In Reference \cite{sym15112001}, the schematic calculations including the quadrupole interaction have been performed for the same. Ref. \cite{sym15112001} shows the effect of the quadrupole interaction of the GT transition strength. They found that the GT strength is highly sensitive to several parameters in the schematic Hamiltonian, and hence, no firm conclusion can be drawn on the shape of the nucleus. The quadrupole term is properly taken into account in the realistic interactions. So, we have compared the experimental GT transition strengths with the realistic interactions.

Recent experimental results on the $\beta$-decay of $^{70}$Kr into $^{70}$Br have been reported in Ref. \cite{VITEZSVEICZER2022137123} together with calculations assuming pseudo-SU (4) symmetry. It is concluded that the new data indicate a restoration of the pseudo-SU(4) symmetry in the $A=70$ nuclei. In this study, we have considered JUN45 \cite{PhysRevC.80.064323}, KB3G \cite{POVES2001157}, GXPF1J \cite{M_Honma_2005} as well as GXPF1A \cite{Honma_2005} effective interactions to calculate the $B_{\rm GT}$ distribution. Corresponding results are depicted in Fig. \ref{GT_Br}. From the recent experiment, the observed $B_{\rm GT}$ value for the first  $1^+$ state is 0.26(3). By comparing the results of GT strength including only pairing matrix elements with the realistic one (presented in Table XXX of Supplemental Material \cite{SM}), we notice that while the dominant contribution with the SDI is due to $f_{5/2}$ $\rightarrow$ $f_{5/2}$, it is the transition of  $g_{9/2}$ to $g_{9/2}$ which contributes dominantly to the total $B_{\rm GT}$ in the JUN45 interaction for this state. Also, the $B_{\rm GT}$ value is increased in the case of the schematic interaction with the $f_{5/2}pg_{9/2}$ model space as compared to the results with the $f_{5/2}p$ model space due to the $g_{9/2}$ $\rightarrow$ $g_{9/2}$ contribution. If we remove the $g_{9/2}$ orbital and include the lower $f_{7/2}$ orbital, the $B_{\rm GT}$ value is reduced significantly. The contribution from $f_{7/2}$ is rather small, indicating the fully occupied $f_{7/2}$ orbital does not contribute to $B_{\rm GT}$ for these nuclei. The 1$^+_{5}$ state has a large $B_{\rm GT}$ value of 0.2816 because of the dominant contribution from $f_{5/2}$ to $f_{5/2}$ for the GXPF1J interaction, similar to the case with the SDI. The cumulative $B_{\rm GT}$ results obtained with the GXPF1A, GXPF1J, and KB3G interactions are close to each other. 

In Fig. \ref{GT_Br}, the experimental $B_{\rm GT}$ values are shown up to an excitation energy of 3.071 MeV. From Fig. \ref{GT_Br}, one can find that the calculations with the JUN45 interaction tend to overestimate the $B_{\rm GT}$ values, likely due to a systematic overestimation from the contribution of the $g_{9/2}$ orbital. The experimental $B_{\rm GT}$ data fall between the results of GXPF1J$^+$ and GXPF1J$^-$. The total $B_{GT}$ is increased when we increase the $np$ pairing strength. With the reduced $np$ pairing strength, we obtain a reduced transition strength. An overall good agreement between the experimental data and the results obtained with the GXPF1A and GXPF1J interactions is observed. 
A more consistent and precise description of the above $N\sim Z$ nuclei may need to have the model space extended to include both $f_{7/2}$ and $g_{9/2}$ orbitals, which would however make the computation very expensive.

It is noticed that there is an enhancement in the GT transition strength for the $\beta$ decay of $^{70}$Kr in comparison to the decay of $^{62}$Ge \cite{VITEZSVEICZER2022137123}, which was suggested as an indication for enhanced $np$ pairing. 
However, no enhancement is observed when we compare the results of theoretically obtained GT strength using the interaction in $f_{5/2}p$ model space containing only pairing matrix elements for the first $1^+$ state (Fig. \ref{SU4}). While, with the inclusion of the $g_{9/2}$ orbital in schematic calculations (Fig. \ref{SDI_5pg9}), the $B_{\rm GT}$ value increases for the first $1^+$ state when we increase the $np$ pairing strength after a certain value due to the increased contribution of the $g_{9/2}$ $\rightarrow$ $g_{9/2}$ transition. One indeed sees a similar increasing trend with the mass number for the first $1^+$ state in calculations of the $\beta$ decay of $^{70}$Kr as compared to the decay of $^{62}$Ge with the JUN45 interaction which is actually due to the enhanced contribution from the $g_{9/2}$ orbital. However, the cumulative GT transition strength decreases. We know that with increasing mass number, the configurations of the states and the model space change. There are several factors that play an important role in the GT transition strength with an increase of mass number. 

We observe that the calculated $B_{\rm GT}$ value for $1_1^+$ decreases with increasing mass number using the GXPF1J interaction. Those calculations significantly underestimate the few observed $B_{\rm GT}$ values, which cannot be corrected solely by the enhancement of the $np$ pairing strength. It should be pointed out that the total accumulated $B_{\rm GT}$ strength increases for the  GXPF1J interaction.  

\section{Conclusion}
We have presented detailed shell model calculations for nuclei with the mass numbers $A=58,62,66$ and $70$, particularly focusing on the $\beta^+$ decay transitions from $Z=N+2$ nuclei to $Z=N$ nuclei and the role played by the $np$ pairing correlation. First, we have determined the impact of pairing on GT transitions as a function of the strength parameter using the SDI with only pairing matrix elements. Even though the interaction is relatively simple, it gives a precise control of the phases of different components of the wave function and separates contributions from different single-particle orbitals to the total $B_{\rm GT}$. An important conclusion one can draw from those calculations with the same coupling strength is that the $B_{\rm GT}$ between yrast $0^+$ and $1^+$ states does not necessarily increase with increasing $np$ pairing strength. By including the $g_{9/2}$ orbital, it can be enhanced due to the increased contribution of $g_{9/2}$. The same conclusion is obtained for the different coupling strengths for both pairing channels. These schematic calculations are compared with large-scale shell model calculations with realistic interactions where we have considered two shell model spaces to understand the role played by different orbitals on the GT strength. We have also done calculations by varying the single-particle energies and $T=0$ $np$ pairing matrix elements in order to investigate the effect of the pesudo-SU(4) symmetry and the isoscalar $np$ pairing on the decay properties. We found an increment in cumulative GT transition strength with increased $np$ pairing matrix elements. The accumulated $B_{\rm GT}$ strengths are rather well reproduced but the decay to the low-lying states can be sensitive to the contributions from the deep-lying $f_{7/2}$ and higher-lying $g_{9/2}$ orbitals. Calculation using the interaction with an extended model space, containing both $f_{7/2}$ and $g_{9/2}$ orbitals, will be required to provide a better picture.

\section*{ACKNOWLEDGMENTS}
PC acknowledges financial support from the Olle Engkvist Foundation. The computations were enabled by resources provided by PDC at the KTH Royal Institute of Technology.

\bibliography{refs}
\end{document}